\documentclass[9pt,twocolumn,twoside]{pnas-new}
\usepackage{siunitx}
\usepackage{graphicx}
\usepackage{multirow}
\usepackage{subcaption}
\usepackage[final]{pdfpages}

\setboolean{displaywatermark}{false}
\templatetype{pnasbriefreport}

\newcommand{\labelphantom}[1]{%
  \parbox{0pt}{\phantomsubcaption\label{#1}}%
}

\title{Super-linear Scaling Behavior for Electric Vehicle Chargers and Road Map to Addressing the Infrastructure Gap}
\author[a]{Alexius Wadell}
\author[a]{Matthew Guttenberg}
\author[b]{Christopher P. Kempes}
\author[a]{Venkatasubramanian Viswanathan}

\affil[a]{Department of Mechanical Engineering, Carnegie Mellon University, Pittsburgh, Pennsylvania 15213, United States}
\affil[b]{The Santa Fe Institute, Santa Fe, New Mexico 87501, United States}

\leadauthor{Wadell}

\authorcontributions{VV conceived the idea for the project, AW performed the data curation, AW and MG performed the data analyses with input from CK and VV, and all authors jointly analyzed the results and wrote the paper.}
\authordeclaration{MG and VV are inventors on a patent related to electric vehicle charger placement.}
\correspondingauthor{\textsuperscript{2}To whom correspondence should be addressed. E-mail: venkvis\@cmu.edu}


\begin{abstract}
Enabling widespread electric vehicle (EV) adoption requires substantial build-out of charging infrastructure in the coming decade.
We formulate the charging infrastructure needs as a scaling analysis problem and use it to estimate the EV infrastructure needs of the US at a county-level resolution.
Surprisingly, we find that the current EV infrastructure deployment scales super-linearly with population, deviating from the sub-linear scaling of gasoline stations and other infrastructure.
We discuss how this demonstrates the infancy of EV station abundance compared to other mature transportation infrastructures.
By considering the power delivery of existing gasoline stations, and appropriate EV efficiencies, we estimate the EV infrastructure gap at the county level, providing a road map for future EV infrastructure expansion.
Our reliance on scaling analysis allows us to make a unique forecast in this domain.
\end{abstract}

\dates{This manuscript was compiled on \today}
\doi{\url{www.pnas.org/cgi/doi/10.1073/pnas.XXXXXXXXXX}}

\begin{document}

\maketitle
\thispagestyle{firststyle}
\ifthenelse{\boolean{shortarticle}}{\ifthenelse{\boolean{singlecolumn}}{\abscontentformatted}{\abscontent}}{}

\dropcap{C}onsumer interest in Electric Vehicles (EV) has been rising as EVs approach price parity with internal combustion engines (ICE).
However, the lack of sufficient Electric Vehicle Supply Equipment (EVSE), popularly called charging stations, could slow future adoption\cite{shareefReviewStageoftheartCharging2016,ahmedEnablingFastCharging2017}.
Thus identifying the placement of EVSE to meet this growing demand is essential and has been an area of extensive research~\cite{shareefReviewStageoftheartCharging2016,guttenbergINCEPTSSoftwareHighfidelity2021,ahmedEnablingFastCharging2017,weiPersonalVehicleElectrification2021}.
Prior work in the area has relied predominantly on mathematical optimization to maximize captured vehicle flow, minimize economic costs, or account for power grid-related issues~\cite{shareefReviewStageoftheartCharging2016}.
Bottom-up optimization based-methods can precisely place and size EVSE based on the local demand and peculiarities.
However, they can become challenging and computationally prohibitive for larger-scale analysis, as a complex system of trade-offs emerges, which brings into question which factors are feasible to include and what form the loss function should take~\cite{shareefReviewStageoftheartCharging2016}.

On the other hand, a coarse-grained scaling analysis approach provides a tractable way to assess the infrastructure needs for large regions.
In particular, scaling relationships demonstrate how features change with system size and often illustrate that a single set of mechanisms is governing a system~\cite{bettencourtOriginsScalingCities2013,kempes2019scales,taylor2019scalability,westScaleUniversalLaws2017,marquetScalingPowerlawsEcological2005}.
Even when the specific mechanisms are unidentified, the coarse-grained regularities of a system, as captured by the scaling exponents, often provide a powerful foundation for building models, making forecasts, and interpreting dominant trade-offs~\cite{bettencourtOriginsScalingCities2013,bettencourtGrowthInnovationScaling2007,taylor2019scalability,kempes2019scales,leitaoThisScalingNonlinear2016,westScaleUniversalLaws2017,marquetScalingPowerlawsEcological2005}.
Here we formulate the EVSE infrastructure problem in a scaling analysis framework \(\left( Y = Y_0 N^{\beta} \right)\), which connects the EVSE infrastructure needs for a particular region to its population size.
We find that the current EVSE infrastructure follows scaling similar to gas stations; however, the scaling exponent is super-linear (\(\beta > 1\)) while that of gas stations is sub-linear (\(\beta < 1\)).
Sub-linear scaling behavior is well established for infrastructure assets and is motivated by theory~\cite{bettencourtOriginsScalingCities2013,bettencourtGrowthInnovationScaling2007}.
The super-linear exponent of the current EVSE build-out indicates interesting differences.
The appearance of super-linear exponents in human systems has been observed for quantities that rely on social interactions, such as patents or gross domestic product~\cite{bettencourtGrowthInnovationScaling2007}.
The argument here is that as cities increase in size, they densify, leading to increased social interactions per person in settings with less infrastructure requirements.
This paradigm implies that all new infrastructures should be sub-linear.
However, this is not necessarily true for an emerging infrastructure, which may be initially driven by social phenomena and later by standard infrastructural geometry constraints.
Thus, we argue that the emergent behavior of the current EVSE market should initially follow super-linear scaling due to social adoption dynamics but will transition to sub-linear behavior as the infrastructure matures.
Finally, we estimate the EVSE infrastructure gap as the number of additional stations required to match the power delivery of existing gasoline stations.
To our knowledge, scaling analysis has not been used to forecast future infrastructure requirements of emerging technology.
Our application here to EVSEs opens up a variety of future work in scaling theory.

\subsection*{Results}

\begin{figure}[t]
    \centering
    \labelphantom{fig:charger_scale_stations}
    \labelphantom{fig:charger_scale_power}
    \includegraphics[width=\linewidth]{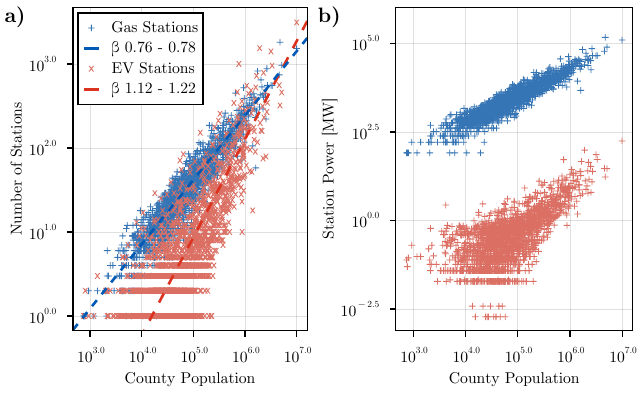}
    \caption{\subref*{fig:charger_scale_stations})~Power Law Scaling for Gas Station and Electric Vehicle Supply Equipment for United State Counties (n = 3143), showing novel super-linear behavior for EVSE stations and expected sub-linear behavior for gas stations. Super-linear behavior suggests EVSE infrastructure has been consolidated in larger population centers. 95\% confidence intervals for the scaling exponent \(\beta\) are shown in the legend. \subref*{fig:charger_scale_power})~Comparison of the power delivery of existing gas stations (Assuming 12 pumps per station and improved efficiency of EVs) to existing EVSE infrastructure. While the EVSE infrastructure of some counties has reached parity, in quantity with gasoline stations, no counties have reached parity in terms of power delivery.\label{fig:charger_scale}
}
\end{figure}

We curated a dataset of county-level EVSE charger, gasoline station counts, and the corresponding county population.
We fit several Generalized Linear Models (GLM) to the data using maximum likelihood estimation (MLE) to predict the number of EVSE (\(Y_{EVSE}\)) and gasoline (\(Y_{GS}\)) stations for all counties (\(n = 3143\)) in the United States from their population (Figure~\ref{fig:charger_scale_stations}).
Using a likelihood ratio test, we compared all models to their null counterparts and have tabulated the test statistic \(\lambda_{LR}\) for each model in Table~\ref{tab:model_zoo}.
All models were found to be highly significant with a p-value of less than \(10^{-99}\), as such, we have only reported the test statistic.

For the power-law scaling model, we initially fit models assuming a Poisson distribution, \(Pois(\mu)\), but found the assumption that the variance is equal to the mean to be a poor fit for our data.
We relaxed this by fitting a Negative Binomial (NB) distribution, \(NB(r, \mu)\) parameterized by a dispersion parameter \(r\) and its expected value \(\mu \).
Using this model, our fitted values for \(\beta \), with a 95\% confidence interval, was \(\beta = 1.17 \pm 0.051\) for the EVSE stations and \(\beta = 0.77 \pm 0.0092\) for the gasoline stations.
We do see a reduction in McFadden's \(R^2_{McF}\) for the NB models~(Table~\ref{tab:model_zoo}) driven by a higher likelihood for their null counterpart, and not a reduction in the model's fit.
We performed a Wald test to determine if \(\beta \neq 1\) was statistically significant for the NB power law models;
and found \(\beta \) to be significantly different from 1 for both the EVSE (\(\mbox{SE} = 0.026\mbox{, W} = 6.4\mbox{, p} < 10^{-9}\)) and gas stations (\(\mbox{SE} = 0.0047\mbox{, W} = -50\mbox{, p} < 10^{-99}\)).
As all models achieved similar Root-Mean-Squared-Deviations (RMSD) and are statically significant (\(\lambda_{LR}\)), we used the Bayesian Information Criterion (BIC) to compare models.
Using the criteria that \(\Delta BIC > 6\) indicates a significant difference in model performance~\cite{leitaoThisScalingNonlinear2016}, we found strong evidence for the NB power scaling models over both linear and quadratic models.
An in-depth discussion of the above statistical tests can be found in the SI appendix.

We performed an ordinary least squares (OLS) regression on the log-log transformed data in line with prior scaling works~\cite{bettencourtOriginsScalingCities2013,bettencourtGrowthInnovationScaling2007,taylor2019scalability,kempes2019scales,leitaoThisScalingNonlinear2016,marquetScalingPowerlawsEcological2005}.
However, as this method does not support zero-count data, it excludes a significant portion of our dataset (43.8\% for EVSE and 1\% for gasoline stations).
We did not find the OLS models compelling and excluded the fits from this analysis; see the SI Appendix for additional details.

\subsubsection*{Gas Station to EVSE Scaling}
Driven by an analogous utility, as EV adoption increases, EVSE infrastructure should tend towards the same coarse-grained regularities as gasoline stations.
With the number of stations in an area to be proportional to the vehicle miles travel \(M\), average efficiency \(E\), and power delivery of a station \(P\): \(Y \propto \eta M / P\).
Assuming no change in consumer driving behavior, the number of EVSE stations to replace a single gasoline station is:

\begin{equation}\label{eq:ev_power_parity}
    \frac{Y_{EVSE}}{Y_{GS}} = \frac{\eta_{EV} P_{GS}}{\eta_{ICE} P_{EVSE}}
\end{equation}

Using the regulated gasoline flow rate of 10 \si{gpm} for a consumer pump in the United States~(40 CFR \S 1090.1550) and the EPA's 33.705\si{kWh/gallon} of gasoline equivalency (40 CFR \S 600.002), the max power delivery of a consumer gasoline pump is \(P_{EVSE} = 20.2 \si{MW}\).
We assume \( \eta_{EV} / \eta_{ICE} \approx 3 \), or that on average EV's consume 1/3 the energy per mile traveled compared to ICE vehicles.

Assuming \(P_{EVSE} = 400\si{kW}\) or Extreme Fast Charging~\cite{ahmedEnablingFastCharging2017}, gives \( Y_{EVSE} / Y_{GS} = 17 \), or to replace one gasoline pump, 17 ports are required to reach power parity.
Reducing \(P_{EVSE}\) to 11.5kW, or the upper end of available home chargers, we find \( Y_{EVSE} / Y_{GS} \approx 586 \).
From this, holding the average number of pumps/ports constant, the number of EVSE stations needed in an area is given by Eq.~\ref{eq:ev_charger_prediction}.

\begin{equation}\label{eq:ev_charger_prediction}
    \hat{Y}_{EVSE} = 3 \frac{P_{GS}}{P_{EV}} \times Y_{0,GS} N^{\beta_{GS}}
\end{equation}

\begin{table}[h]
    \caption{Model Statistics for Fitted Generalized Linear Models\hspace*{\fill}}\label{tab:model_zoo}
    \centering
    \begin{tabular}{llccccc}
 & Model & RMSD & $R^2_{McF}$ & \shortstack{$\lambda_{LR}$ \\ $\times 10^{3}$} & \shortstack{BIC \\ $\times 10^{3}$} \\
\midrule 
\multirow{4}{*}{\rotatebox[origin=c]{90}{EVSE}} & $NB \left (r, \mu = Y_0 N ^\beta \right)$ & $49.5$ & $0.195$ & $3.24$ & $13.4$ \\
 & $\mathcal{N}(a N^2 + bN + c, \sigma^2)$ & $45.3$ & $0.104$ & $3.81$ & $32.9$ \\
 & $\mathcal{N}(a N + b, \sigma^2)$ & $49.6$ & $0.088$ & $3.25$ & $33.5$ \\
 & $Pois(Y_0 N ^\beta)$ & $46.4$ & $0.804$ & $167$ & $40.7$ \\
\midrule 
\multirow{4}{*}{\rotatebox[origin=c]{90}{Gasoline}} & $NB \left (r, \mu = Y_0 N ^\beta \right)$ & $29.2$ & $0.283$ & $8.02$ & $20.3$ \\
 & $Pois(Y_0 N ^\beta)$ & $28.8$ & $0.862$ & $176$ & $28.2$ \\
 & $\mathcal{N}(a N^2 + bN + c, \sigma^2)$ & $28.6$ & $0.168$ & $6.08$ & $30.0$ \\
 & $\mathcal{N}(a N + b, \sigma^2)$ & $30.6$ & $0.156$ & $5.63$ & $30.5$ \\
\bottomrule 
\end{tabular}

\end{table}

\subsubsection*{Mean-field Model Relating Charging Power and Vehicle Speed}

In this section, we develop a mean-field model to estimate the maximum average speed of an EV for a given charger power.
We assume that the EV charges at \(P_{EVSE}\) for \(\alpha \) percent of the time, then drives at a speed of \(v\) for the remaining \(1 - \alpha \) percent of the time.
We further make the simplifying assumption that aerodynamic forces dominate the overall power draw of the EV.\@
For a continuous drive cycle, the energy delivered during charging must match the energy demand of the drive cycle:

\begin{equation}
    P_D (1 - \alpha) = \alpha P_{EVSE} \to
    \frac{1}{2} \rho C_d A v^3 = \frac{\alpha}{1 - \alpha} P_{EVSE}
\end{equation}

From this, we can relate the power of the charger \(P_{EVSE}\) to the vehicle's speed while driving \(v\):

\begin{equation}\label{eq:home_charging_speed}
    v = \sqrt[3]{ \frac{\alpha}{1-\alpha} \frac{2}{\rho C_d A} P_{EVSE} }
\end{equation}

As the vehicle is stationary while charging, the average vehicle speed is \(\bar{v} = v (1-\alpha) \propto \alpha^{1/3} {\left( 1-\alpha \right)}^{2/3} \).
The maximum average speed occurs when the vehicle spends \(\alpha = 1/3\) of its time charging at 2x the power it consumes while driving~(Figure~\ref{fig:home_chargers_avg_speed}).
For our analysis we assume \( \rho = 1.225 \si{kg/m^3}\) and \(C_d A = 0.75 \si{m^2} \); or a boxy sedan at International Standard Metric Conditions.
In the case of \(P_{EVSE} = 1.92 \si{kW}\), this translates to a driving speed of \(v = 28.6 \si{mph}\), and an average speed of \( \bar{v} = 19 \si{mph} \), speeds well suited for residential driving.
Higher driving speeds are possible with 1.92\si{kW}, but only at the expense of increased charging times, which may be tenable for sort trips if charging can occur before the next trip.
However, operating above \(\alpha > 1/3\) will increase the overall trip duration for long-range trips.
At the other extreme, 400 \si{kW} delivers a max driving speed (\textasciitilde170mph) far beyond highway speed limits.
Instead, the higher charging rates enable shorter charging times for a given driving speed (Figure~\ref{fig:home_chargers_driving_speed}).

Our assumption that aerodynamic forces dominate the overall power consumption breaks down at low speeds.
More complex models are possible\cite{guttenbergINCEPTSSoftwareHighfidelity2021,weiPersonalVehicleElectrification2021}, but do not provide a closed-form solution for velocity as a function of charger power.
Additionally, as a lower bound on the vehicle's power consumption, Eq.~\ref{eq:home_charging_speed} provides an upper bound on the vehicle's speed for a given charger power.
Thus the qualitative conclusion that charger power limits the types of trips that are feasible remains valid.

\subsubsection*{Home Charging}

At present, consumer-grade chargers range from 1.92kW (120V@12A) to 11.5kW (240V@48A), significantly slower than commercial or a ``gasoline-equivalent'' charger.
As shown in Figure~\ref{fig:home_chargers_avg_speed}, home chargers are insufficient for a continuous operating cycle at highway speeds.
The mean-field analysis provides a framework to understand the scaling relationship between EVSE power and vehicle speed.
A more detailed analysis is needed to account for real-world factors such as mixed power level charging stations and idle time for the vehicle.
Home charging may be sufficient for many consumers' local commutes~\cite{weiPersonalVehicleElectrification2021}; however, it will be insufficient for predominantly long-distance highway travel.

\begin{figure}[t]
    \centering
    \labelphantom{fig:home_chargers_avg_speed}
    \labelphantom{fig:home_chargers_driving_speed}
    \includegraphics[width=\linewidth]{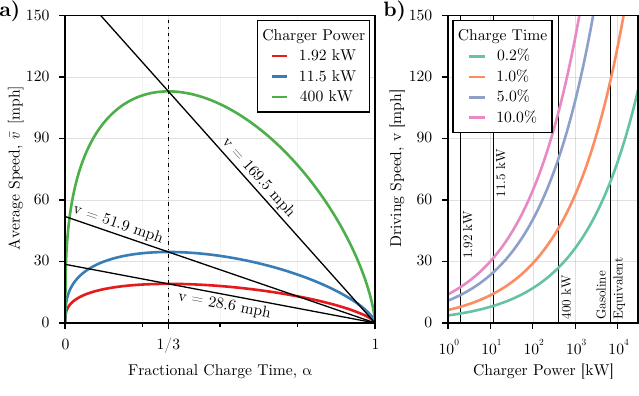}
    \caption{\subref*{fig:home_chargers_avg_speed})~Average vehicle speed vs.~time charging for varying charger power outputs. The solid black lines of constant driving speed (\(v\)) reflect the impact of increasing charge time, when the vehicle is stationary, on the average speed (\(\bar{v}\)). The maximum \(\bar{v}\) for a given charger power occurs at \(\alpha = 1/3\), indicated by the dashed vertical bar. \subref*{fig:home_chargers_driving_speed})~Trade-off between driving speed \(v\) and charger power for various charging times, with vertical lines at notable charger power levels. For example, a vehicle driving at 30mph would spend \textasciitilde10\% of it's time charging at 11.5kW vs.\ \textasciitilde0.2\% at 400kW.\label{fig:home_chargers}}
\end{figure}

\subsection*{EVSE Infrastructure Gap}
\begin{figure}
    \centering
    \includegraphics[width=\linewidth]{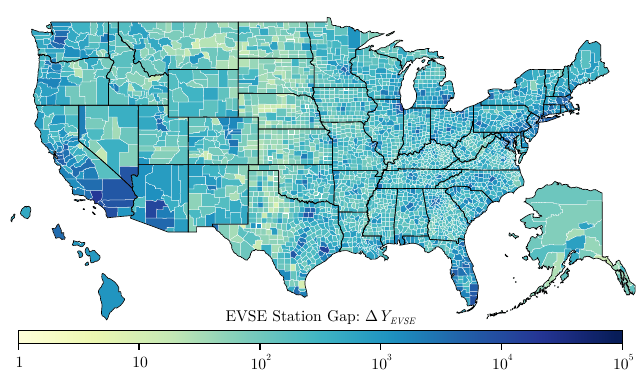}
    \caption{The EVSE Station Gap between the number of existing EVSE stations and the number predicted with Eq.~\ref{eq:ev_charger_prediction}, assuming all current and future chargers are capable of 400kW.
    At present, no counties have sufficient EVSE stations to meet power parity, even when assuming all existing stations have been upgraded to 400 kW.
    }\label{fig:us_charger_gap}
\end{figure}

Using Eq.~\ref{eq:ev_charger_prediction}, we estimated the EVSE Station Gap: \(\Delta Y_{EVSE} = \hat{Y}_{EVSE} - Y_{EVSE}\), or the number of additional stations needed for each county, assuming all current and future chargers are 400kW~(Figure~\ref{fig:us_charger_gap}).
The gap between the existing EVSE infrastructure and the existing ICE infrastructure is large (Figure~\ref{fig:charger_scale_power}).

Our model assumes a fixed average pumps/ports ratio, ignores consumer behavior regarding longer charging times, the role of home chargers, and variations in consumer behavior over time.
Our treatment of existing gasoline stations neglect factors, such as gasoline purchases outside of on-highway transportation, that may inflate the number of stations in a region.
Finally, scaling laws can not provide precise EVSE placement or sizing information as a coarse-grained model.
Further refinement is left to optimization-based methods~\cite{shareefReviewStageoftheartCharging2016} or complex systems analysis~\cite{guttenbergINCEPTSSoftwareHighfidelity2021}.

\subsection*{Temporal Evolution of Scaling Relationships}

Beyond the planning and policy implications of the scaling relationships presented here, these findings also provide new insights into scaling phenomena in general.
The central question in scaling theory is how to connect scaling exponents with fundamental mechanisms.
For many biological systems and human infrastructure, the optimal solution to dominant physical constraints directs the scaling exponent~\cite{kempes2019scales,bettencourtOriginsScalingCities2013,westScaleUniversalLaws2017,marquetScalingPowerlawsEcological2005}.
At the same time, recent studies show that human institutions can adjust their exponent values based on their distinct goals and missions~\cite{taylor2019scalability}.
However, in biology, we have not observed the evolution of scaling exponent values in time towards the equilibrium optimum.
Such temporal changes would have occurred during much deeper histories than we can observe.
EVSE infrastructure is an example of an out-of-equilibrium scaling relationship, where predicting the equilibrium scaling exponent provides unique planning forecasts.

\matmethods{
County-level population estimates were obtained from the United States Census Bureau's Population Estimates Program using the 2020 Vintage estimates of the 2010 Census; at the time of writing, this was the most up-to-date estimate available~\cite{CountyPopulationTotalsC2010V2020}.
We used the Forth Quarter 2020 counts for ``Gasoline Stations'' from the United States Bureau of Labor Statistics Quarterly Census of Employment and Wages Program~\cite{QuarterlyCensusEmployment2021}.
EVSE charger locations, and other metadata, were obtained from the National Renewable Laboratory's Alternative Fuel Stations Application Programming Interface~\cite{nrelAlternativeFuelStations2021} and then geocoded to counties using shapefiles obtained from the United States Census~\cite{census2020TIGERLine2021}.
We then fit various GLMs to the data to predict station counts, EVSE and Gasoline, for each county using maximum likelihood estimation as implemented in the Julia Package GLM.jl.
Due to difficulty fitting some parameters, we increased the maximum iterations from 30 to 60 but otherwise used the default settings.
For an extended discussion of model fitting, see the GLM section of the SI Appendix.
Total EVSC Power for each station was computed by multiplying the counts of Level 1 (1.4\si{kW}), Level 2 (7.2\si{kW}), and DC Fast Chargers (50\si{kW}) by their respective power outputs shown in parenthesis~\cite{ahmedEnablingFastCharging2017}.
Individual EVSE station power estimates were summed to produce the county-level EVSE power estimates shown in Figure~\ref{fig:charger_scale_power}.

}
\showmatmethods{} 

\acknow{
    The authors from CMU acknowledge the support from Mobility21, A United States Department of Transportation National University Transportation Center.
    CPK thanks CAF Canada and Toby Shannan for generously supporting this work.
}

\showacknow{} 

\bibliography{references}

\begin{thebibliography}{10}

\bibitem{shareefReviewStageoftheartCharging2016}
H Shareef, MM Islam, A Mohamed, A review of the stage-of-the-art charging
  technologies, placement methodologies, and impacts of electric vehicles.
\newblock {\em\protect\JournalTitle{Renewable and Sustainable Energy Reviews}}
  \textbf{64}, 403--420 (2016).

\bibitem{ahmedEnablingFastCharging2017}
S Ahmed, et~al., Enabling fast charging \textendash{} {{A}} battery technology
  gap assessment.
\newblock {\em\protect\JournalTitle{Journal of Power Sources}} \textbf{367},
  250--262 (2017).

\bibitem{guttenbergINCEPTSSoftwareHighfidelity2021}
M Guttenberg, S Sripad, A Bills, V Viswanathan, {{INCEPTS}}: {{Software}} for
  high-fidelity electric vehicle en route state of charge estimation, fleet
  analysis and charger deployment.
\newblock {\em\protect\JournalTitle{eTransportation}} \textbf{7}, 100106
  (2021).

\bibitem{weiPersonalVehicleElectrification2021}
W Wei, S Ramakrishnan, ZA Needell, JE Trancik, Personal vehicle electrification
  and charging solutions for high-energy days.
\newblock {\em\protect\JournalTitle{Nature Energy}} \textbf{6}, 105--114
  (2021).

\bibitem{bettencourtOriginsScalingCities2013}
LMA Bettencourt, The {{Origins}} of {{Scaling}} in {{Cities}}.
\newblock {\em\protect\JournalTitle{Science}} \textbf{340}, 1438--1441 (2013).

\bibitem{kempes2019scales}
CP Kempes, M Koehl, GB West, The scales that limit: The physical boundaries of
  evolution.
\newblock {\em\protect\JournalTitle{Frontiers in Ecology and Evolution}}
  \textbf{7}, 242 (2019).

\bibitem{taylor2019scalability}
RC Taylor, et~al., The scalability, efficiency and complexity of universities
  and colleges: {{A}} new lens for assessing the higher educational system.
\newblock {\em\protect\JournalTitle{arXiv preprint arXiv:1910.05470}} (2019).

\bibitem{westScaleUniversalLaws2017}
GB West, {\em Scale: The Universal Laws of Growth, Innovation, Sustainability,
  and the Pace of Life in Organisms, Cities, Economies, and Companies}.
\newblock ({Penguin Press}, {New York}), (2017).

\bibitem{marquetScalingPowerlawsEcological2005}
PA Marquet, et~al., Scaling and power-laws in ecological systems.
\newblock {\em\protect\JournalTitle{Journal of Experimental Biology}}
  \textbf{208}, 1749--1769 (2005).

\bibitem{bettencourtGrowthInnovationScaling2007}
LMA Bettencourt, J Lobo, D Helbing, C K{\"u}hnert, GB West, Growth, innovation,
  scaling, and the pace of life in cities.
\newblock {\em\protect\JournalTitle{Proceedings of the National Academy of
  Sciences}} \textbf{104}, 7301--7306 (2007).

\bibitem{leitaoThisScalingNonlinear2016}
JC Leitao, JM Miotto, M Gerlach, EG Altmann, Is this scaling nonlinear?
\newblock {\em\protect\JournalTitle{Royal Society Open Science}} \textbf{3},
  150649 (2016).

\bibitem{CountyPopulationTotalsC2010V2020}
{US Census Bureau}, County {{Population Totals}}: 2010-2020 (2021).

\bibitem{QuarterlyCensusEmployment2021}
{U.S. Bureau of Labor Statistics}, Quarterly {{Census}} of {{Employment}} and
  {{Wages}} (2021).

\bibitem{nrelAlternativeFuelStations2021}
{National Renewable Energy Laboratory}, Alternative {{Fuel Stations API}}
  (2021).

\bibitem{census2020TIGERLine2021}
{U.S. Census Bureau}, 2020 {{TIGER}}/{{Line Shapefiles}} for {{County}} and
  {{Equivalent}} (2021).

\end{thebibliography}

\clearpage


\section*{Supplementary material}

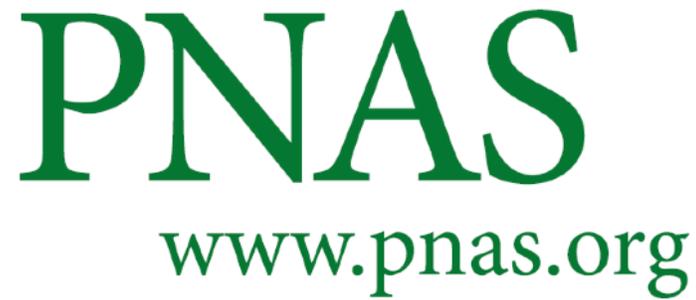

# Supplementary Information for

## Super-linear Scaling Behavior for Electric Vehicle Chargers and Roadmap to Addressing the Infrastructure Gap

Alexius Wadell, Matthew Guttenberg, Christopher P. Kempes, and Venkatasubramanian Viswanathan

Venkatasubramanian Viswanathan.
E-mail: venkvis@cmu.edu

**This PDF file includes:**

SI Appendix
SI References

## Supporting Information Appendix

### 1. Generalized Linear Models

Generalized Linear Models (GLM), extend linear regression by introducing a non-linear link function $g$ such that the expected response is: $E[Y|X] = g^{-1}(\eta(X))$, where $\eta(X)$ is the linear function of the dependent variables $X$. Additionally, GLMs allow distribution of $Y$ to be any distribution $f(Y_i; \theta_i = g(\mu), \phi_i)$ from the exponential family, where $\theta_i$ is the natural parameter of the distribution, and $\phi_i$ is the dispersion parameter of the distribution. We use $i = 1 \ldots n$ to index over our $n$ observations, and related quantities.

Table 1 of the paper tabulated the various models $Y \sim f\left(\mu = g^{-1}(\eta(N))\right)$ that we examined. For the power law models, we used the following for $g$ and $\eta(N)$:

$$g(\mu) = \ln \mu$$
$$\eta(N) = \ln Y_0 + \beta \ln N \qquad [1]$$

As the expected value of the model $E[Y|N] = g^{-1}(\eta(N))$, this parameterization results in the desired power law model:

$$\mu(N) = g^{-1}\left(\eta(N)\right) \to \exp\left(\ln Y_0 + \beta \ln N\right) \to Y_0 N^\beta$$

**A. Maximum Likelihood Estimation.** Maximum likelihood estimation (MLE) is a method for estimating the parameters of a model that are most probable given the observed data. This results in an unbiased estimate of the model parameters, that are the most probable values of the parameters, and in the limit of large sample sizes, approach the true values of the parameters.

For a GLM, the likelihood of the model $\mathcal{L}$ is defined by Eq. 2, where $f$ is the probability distribution of the model.

$$\mathcal{L} = \prod_i^n f(Y_i; \theta_i, \phi) \qquad [2]$$

$$f(Y_i; \theta_i, \phi) = \exp\left(\frac{Y_i \theta_i - b(\theta_i)}{a(\phi)} + c(Y_i, \phi)\right) \qquad [3]$$

For numerical stability, we maximize the log-likelihood $\ell$ of the model, rather than the likelihood $\mathcal{L}$. Finding the parameters that maximize Eq. 4 was handled by GLM.jl (1) using the Iteratively Reweighed Least Squares algorithm (2). The implementation details of the algorithm are beyond the scope of this document.

$$\max \sum_i^n \ln f(Y_i; \theta_i, \phi) \to \max \sum_i^n \ell_i \qquad [4]$$

The log-likelihood of $f$ for a single observation $Y_i$ is given by:

$$\ell_i = \ln f(Y_i; \theta_i, \phi) \to \frac{Y_i \theta_i - b(\theta_i)}{a(\phi)} + c(Y_i, \phi) \qquad [5]$$

The expected value $E[Y_i] = \mu_i$, occurs when $E[\partial \ell_i / \partial \theta_i] = 0$, thus $\mu_i = b'(\theta_i)$, where $b'$ is the first derivative of $b$.

$$0 = E\left[\frac{\partial \ell_i}{\partial \theta_i}\right] \to E\left[\frac{Y_i - b(\theta_i)}{a(\phi)}\right] \to \frac{E\left[Y_i - b'(\theta_i)\right]}{a(\phi)} \to E[Y_i] = b'(\theta_i) \qquad [6]$$

The canonical link function $g$ of a distribution $f$ is the inverse of $b'$, and thus transforms $\theta$ to the mean of the distribution. Alternative link functions are permissible, but the canonical link enables further simplification of the likelihood function (3).

**A.1. Poisson Distribution.** For the Poisson power law models $Y \sim Pois(Y_0 N^\beta)$, we used the following for $g(\mu)$ and $\eta(N)$:

$$g(\mu) = \ln \mu$$
$$\eta(N) = \ln Y_0 + \beta \ln N$$

Substituting into Eq. 5, for a single observation we get:

$$\ln f(Y_i; \theta_i) = \ln \frac{\theta_i^{Y_i} e^{-\theta_i}}{Y_i!} \to Y_i \ln \theta_i - \theta_i - \ln(Y_i!) \to Y_i \cdot \eta(N) - \exp(\eta(N)) - \ln(Y_i!)$$

This gives the criterion for the maximum likelihood estimate for the poisson power-law models:

$$\operatorname*{argmax}_{Y_0, \beta} \sum_i^n Y_i(\ln Y_0 + \beta \ln N_i) - Y_0 N_i^\beta$$

Alexius Wadell, Matthew Guttenberg, Christopher P. Kempes, and Venkatasubramanian Viswanathan

**A.2. Negative Binomial Distribution.** The negative binomial distribution $NB(Y_i; \theta_i, r)$ has the probability density function given by Eq. 7, where $r$ is a shape parameter, and $\theta_i$ is the expected value of the distribution.

$$NB(Y_i; \theta_i, r) = \frac{\theta^{Y_i}}{Y_i!} \frac{\Gamma(Y_i + r)}{\Gamma(r)(\theta_i + r)^{Y_i}} \frac{1}{(1 + \theta_i/r)^r} \quad [7]$$

This is a transformation of the $NB(Y_i; r, p)$ parameterization, using $\theta = pr/(1-p)$ (4), such that $E[Y] = \theta$ and the variance is $\text{Var}[Y] = \theta(1 + \theta/r)$. Additionally, in the limit of $r \to \infty$, the negative binomial distribution is equivalent to the Poisson distribution. Thus it allows us to extend the Poisson distribution to account for over-dispersion, instead of enforcing $\text{Var}(Y) = E[Y]$ (4, 5).

The negative-binomial power-law models $Y \sim NB(\mu = Y_0 N^\beta, r)$ using Eq. 1 for the $g(\mu)$ and $\eta(N)$. Substituting Eq. 7, into Eq. 5 gives the log-likelihood for a single observation:

$$\ln NB(Y_i; \theta_i, \theta) = Y_i \ln \theta_i - Y_i \ln(\theta_i + r) - r \ln(1 + \theta_i/r) - \ln(Y_i!) + \ln \Gamma(Y_i + r) - \ln \Gamma(r)$$

$$= Y_i(\ln Y_0 + \beta \ln N_i) - Y_i \ln \left(Y_0 N_i^\beta + r\right) - r \ln\left(1 + \frac{Y_0 N_i^\beta}{r}\right) - \ln(Y_i!) + \ln \Gamma(Y_i + r) - \ln \Gamma(r)$$

Plugging in the above equation to Eq. 4, and dropping terms that are independent of the parameters ($Y_0$, $\beta$ and $r$), gives the following criterion for the maximum likelihood estimate:

$$\operatorname*{argmax}_{Y_0, \beta, r} \sum_i^n \left[ Y_i(\ln Y_0 + \beta \ln N_i) - Y_i \ln \left(Y_0 N_i^\beta + r\right) - r \ln\left(1 + \frac{Y_0 N_i^\beta}{r}\right) \right]$$

As the negative binomial is only a GLM for fixed $r$, we fit the model using Expectation-Maximization to iteratively alternate between finding MLE fits for the model parameters ($Y_0$ and $\beta$) and $r$ (1); until we converge to an MLE estimate for all three parameters. This procedure is consistent with other statistical packages(5), and results in a MLE estimate for both the model parameters and the dispersion parameter $r$ (4).

## 2. Ordinary Least Squares

Prior works in scaling analysis have minimized the least-squares criterion of the log-transformed data (6–9), as shown in Eq. 8. Expanding Eq. 8 by dropping terms that are independent of $Y_0$ and $\beta$ as well as transforming to a maximization problem results in Eq. 9, and is equivalent to a GLM model of the form: $\ln Y \sim \mathcal{N}(\ln Y_0 + \beta \ln N, \sigma^2)$.

$$\operatorname*{argmin}_{Y_0, \beta} \sum_i^n \left[\ln Y_i - (\ln Y_0 + \beta \ln N_i)\right]^2 \to \sum_i^n \left[\ln(Y_i)^2 - 2 \ln Y_i (\ln Y_0 + \beta \ln N_i) + (\ln Y_0 + \beta \ln N_i)^2\right] \quad [8]$$

$$\operatorname*{argmax}_{Y_0, \beta} \frac{1}{2} \sum_i^n \left[\ln Y_i \cdot (\ln Y_0 + \beta \ln N_i) - \frac{1}{2}(\ln Y_0 + \beta \ln N_i)^2\right] \quad [9]$$

As noted by Li et al., this is an MLE estimate of the model assuming the fluctuations between $\ln Y$ and $\ln Y_0 + \beta$ are normally distributed (10), and the probability of zero-count data is zero: $P(Y = 0) = 0$. However, as 43.8% of counties in our dataset have no EVSE stations, $P(y = 0) = 0$ is not a reasonable approximation. For comparison purposes only, we have computed OLS fits of the log-transformed data for both EVSE and gasoline stations but have excluded them from our analysis in favor of models that can fully explain the data.

For the gasoline station, our fit had a $R^2_{McF} = 0.667$, a likelihood ratio of $\lambda_{LR} = 6.67 \cdot 10^3$ and a BIC score of $3.36 \cdot 10^3$ based on 3,111 out of 3,143 counties. The fitted scaling exponent was $\beta = 0.77 \pm 0.0099$, shows close agreement with previous work(6), and the fitted exponent for the NB model ($\beta = 0.77 \pm 0.0092$). These results suggest that approximating the power-law fit with log-transformed data is reasonable for gasoline stations.

For the EVSE stations, our fit had a $R^2_{McF} = 0.246$, a likelihood ratio of $\lambda_{LR} = 1.57 \cdot 10^3$ and a BIC score of $33.0 \cdot 10^3$, based on 1,765 out of 3,143 counties. The fitted scaling exponent was $\beta = 0.83 \pm 0.032$, in contrast to the $\beta = 1.17 \pm 0.051$ predicted by the NB power model. These results suggest that excluding zero-count data significantly alters the model's fit and is inappropriate for the EVSE dataset. It should also be noted that even with the OLS fits the EVSE scaling exponent is significantly larger than the exponent for gasoline stations.

## 3. Statistical Testing

In order to check the quality of the various GLM fits, perform model selection, and check the significance of the fitted parameters, we performed a variety of statistical tests.

**A. McFadden's pseudo-$R^2$.** $R^2_{McF}$ compares the log-likelihood of the model ($\ln \mathcal{L}$) to the log-likelihood of the null model ($\ln \mathcal{L}$).

$$R^2_{McF} = 1 - \frac{\ln \mathcal{L}}{\ln \mathcal{L}_0} \qquad [10]$$

To compute $\ln \mathcal{L}_0$, we fit a null model using the same link and distribution, but with a constant $\eta$ and use it's log-likelihood for $\ln \mathcal{L}_0$. As noted by McFadden, $R^2_{McF}$ is typically smaller than the $R^2$ of linear regression, and value of 0.2 to 0.4 represents an "excellent fit"(11). Additionally, as $R^2_{McF}$ is a comparison of a model to its null model, it can be used to compare disparate models.

**B. Root Mean Squared Deviation.** The Root Mean Squared Deviation (RMSD), or Root Mean Squared Error (RMSE) is defined as:

$$\text{RMSD} = \sqrt{\frac{1}{n} \sum_{i}^{n} \left(\hat{Y}_i - Y_i\right)^2}$$

Where $\hat{Y}_i$ are the predictions generated by the fitted GLM, $Y_i$ are the observed values, $n$ is the number of observations, and $i = 1 \ldots n$ indexes the observations.

**C. Likelihood Ratio Test.** The likelihood ratio test can be used to check if the fitted model is significantly better than the null model. For large sample sizes, test statistic $\lambda_{LR}$ is chi-squared ($\mathcal{X}^2$) distributed under the null hypothesis that the model is not significantly better than the null model (12). With the alternative hypothesis, that model's fit is significantly different from the null model.

$$\lambda_{LR} = -2 \left(\ln \mathcal{L}_0 - \ln \mathcal{L}\right)$$

We then compare the test statistic to the critical value for $\mathcal{X}^2_k > \lambda_{LR}$ where $k$, the degrees of freedom for $\mathcal{X}^2$, is the difference in the degrees of freedom between the model and the null model. We found all models to be highly significant with $p < 10^{-99}$ for all models, as such, we have only reported the test statistic $\lambda_{LR}$, and not their respective p-values as well. To compute $\ln \mathcal{L}_0$, we fit a null model using the same link and distribution, but with a constant $\eta$ and use it's log-likelihood for $\ln \mathcal{L}_0$.

**D. Parameter Significance.** To check if the fitted parameters of each GLM model are significantly different from zero, we performed a Wald test to check if the model parameters are significantly different from zero(13).

$$W = \frac{\bar{X} - \mu}{SE}$$

Where $\bar{X}$ is the value of the fitted parameter, $\mu$ is its value under the null hypothesis, and $SE$ is the standard error of the parameter as computed while fitting the model. We then compare the test statistic to the critical value that $W < |\mathcal{N}(0, 1)|$. For all fitted models, we found nearly all parameters to be significantly different from zero ($p < 10^{-15}$). For the quadratic model for EVSE count data, we failed to reject the null hypothesis that the intercept parameter was significantly different from zero $p < 0.1$. As the other parameters were significantly different from zero, this does not represent a degenerate model.

In addition, for the power-law fits, we repeated the test to check if $\beta$ was significantly different from ones(13). As noted in the main paper, we found $\beta$ to be significantly different from 1 for both the gasoline stations (SE = 0.026, W = 6.4, p $< 10^{-9}$), and EVSE stations (SE = 0.0047, W = $-50$, p $< 10^{-99}$).

**E. Bayesian Information Criteria.** To compare the predictive power of various models, we computed the Bayesian Information Criteria (BIC) for each of the fitted GLMs. The BIC is based on the log-likelihood ($\ln \mathcal{L}$), the number of observations ($n$) and the number of parameters ($k$), or complexity, of the model (14).

$$\text{BIC} = k \ln n - 2 \ln \mathcal{L} \qquad [11]$$

We can then compare two models by using the BIC scores to estimate the Bayes' Factor $B_{12}$ that one model is better than the other (10, 15).

$$B_{12} \approx \exp\left(\frac{1}{2} \Delta BIC\right) \qquad [12]$$

Where $\Delta \text{BIC} = \text{BIC}_2 - \text{BIC}_1$, thus if $\text{BIC}_1 < \text{BIC}_2$ then model 1 is better than model 2; and the likelihood ratio of model 1 vs. 2 is given by Eq. 12. As proposed by Lei et al., we have used $\Delta \text{BIC} > 6$ as our threshold to declare one model is significantly better than the other. This corresponds to a Bayes' Factor of $B_{12} \approx 20.1$ or that one model is at least 20 more likely to describe the data than the other.

Alexius Wadell, Matthew Guttenberg, Christopher P. Kempes, and Venkatasubramanian Viswanathan

| | | |
|---|---|---|
| Manuscript Title: | Super-linear Scaling Behavior for Electric Vehicle Chargers and Roadmap to Addressing the Infrastructure Gap | |
| Submitting Author*: | Alexius Wadell | |
| # | Question | Y/N/NA[†] |
| 1 | Have you provided all assumptions, theory, governing equations, initial and boundary conditions, material properties, e.g., open circuit potential (with appropriate precision and literature sources), constant states, e.g., temperature, etc.? **Remarks:** Yes, we have provided an in-depth overview of the Generalized Linear Models (GLM) that we used to fit power scaling relations, as well as, the derivation behind our mean-field analysis of home charging. We have documented all constants (Drag Area, Pumps Per Station, Efficiency of EVs vs ICE) and have provided justifications for their inclusion. | Y |
| 2 | If the calculations have a probabilistic component (e.g. Monte Carlo, initial configuration in Molecular Dynamics, etc.), did you provide statistics (mean, standard deviation, confidence interval, etc.) from multiple ($\geq 3$) runs of a representative case? **Remarks:** Calculations, as presented, do not have a probabilistic component. We have provided statistics for our fitted GLM models, however we used the mean expected value for all subsequent calculations. | NA |
| 3 | If data-driven calculations are performed (e.g. Machine Learning), did you specify dataset origin, the rationale behind choosing it, what all information does it contain and the specific portion of it being utilized? Have you described the thought process for choosing a specific modeling paradigm? **Remarks:** We rely on datasets provided by the National Renewable Energy Laboratory, United States Census Bureau and the United States Bureau of Labor Statistics for the data used to fit our power scaling laws. We have documented this in the main document, and will provide scripts for harvesting the requisite data from the above sources. | Y |
| 4 | Have you discussed all sources of potential uncertainty, variability, and errors in the modeling results and their impact on quantitative results and qualitative trends? Have you discussed the sensitivity of modeling (and numerical) inputs such as material properties, time step, domain size, neural network architecture, etc. where they are variable or uncertain? **Remarks:** We have provided a thorough enumeration of the possible sources of error in the main text. As a projection of the future infrastructure needs of the United States, our analysis is inherently speculative. However, we have quantified possible sources of uncertainty and explained how our projections could be modified to account for different hypothetical scenarios (ie. Different EVCS power levels). We performed | Y |
| 5 | Have you sufficiently discussed new or not widely familiar terminology and descriptors for clarity? Did you use these terms in their appropriate context to avoid misinterpretation? Enumerate these terms in the 'Remarks'. **Remarks:** As we sit at the intersection of a few fields (Urban Scaling Analysis, EVCS Infrastrucure Modeling) some terminology is likely to be unfamiliar to our readers. We have defined all terms in the main text as the relevant and have provided citations to relevant literature for further exploration. All terms were used in their appropriate context to the best of our understanding. In the Main Text: Electric Vehicle Charging Station (EVCS), Electric Vehicle (EV), Internal Combustion Engines (ICE). Expanded upon in the SI: Negative Binomial, Poisson, Bayesian Information Criteria, Generalized Linear Models | Y |

* I verify that this form is completed accurately in agreement with all co-authors, to the best of my knowledge.

[†] Y ≡ the question is answered completely. Discuss any N or NA response in 'Remarks'. Alexius Wadell, Matthew Guttenberg, Christopher P. Kempes, and Venkatasubramanian Viswanathan 1 of 1 .

## References

1. D Bates, et al., JuliaStats/GLM.jl: V1.5.1 (Zenodo) (2021).
2. PW Holland, RE Welsch, Robust regression using iteratively reweighted least-squares. *Commun. Stat. - Theory Methods* **6**, 813–827 (1977).
3. A Agresti, *Foundations of Linear and Generalized Linear Models*, Wiley Series in Probability and Statistics. (John Wiley & Sons Inc, Hoboken, New Jersey), (2015).
4. JF Lawless, Negative binomial and mixed poisson regression. *Can. J. Stat.* **15**, 209–225 (1987).
5. A Zeileis, C Kleiber, S Jackman, Regression Models for Count Data in *R*. *J. Stat. Softw.* **27** (2008).
6. LMA Bettencourt, J Lobo, D Helbing, C Kühnert, GB West, Growth, innovation, scaling, and the pace of life in cities. *Proc. Natl. Acad. Sci.* **104**, 7301–7306 (2007).
7. LMA Bettencourt, The Origins of Scaling in Cities. *Science* **340**, 1438–1441 (2013).
8. C Cottineau, O Finance, E Hatna, E Arcaute, M Batty, Defining urban clusters to detect agglomeration economies. *Environ. Plan. B: Urban Anal. City Sci.* **46**, 1611–1626 (2019).
9. PA Marquet, et al., Scaling and power-laws in ecological systems. *J. Exp. Biol.* **208**, 1749–1769 (2005).
10. JC Leitao, JM Miotto, M Gerlach, EG Altmann, Is this scaling nonlinear? *Royal Soc. Open Sci.* **3**, 150649 (2016).
11. D McFadden, Quantitative Methods for Analyzing Travel Behaviour on Individuals: Some Recent Developments, (Cowles Foundation for Research in Economics, Yale University), Cowles Foundation Discussion Papers (1977).
12. SS Wilks, The Large-Sample Distribution of the Likelihood Ratio for Testing Composite Hypotheses. *The Annals Math. Stat.* **9**, 60–62 (1938).
13. L Wasserman, *All of Statistics: A Concise Course in Statistical Inference*, Springer Texts in Statistics. (Springer New York, New York, NY), (2004).
14. G Schwarz, Estimating the Dimension of a Model. *The Annals Stat.* **6** (1978).
15. T Hastie, R Tibshirani, J Friedman, *The Elements of Statistical Learning*, Springer Series in Statistics. (Springer New York, New York, NY), Second edition, (2009).



\end{document}